\documentstyle[epsfig]{aipproc}
\newcommand{\beq}{\begin{equation}}
\newcommand{\eeq}{\end{equation}}
\newcommand{\bea}{\begin{eqnarray}}
\newcommand{\eea}{\end{eqnarray}}
\newcommand{\apj}{{\it Astrophys. J.} }
\begin{document}
\title{Self-Similar Hot Accretion Flow onto a Neutron Star}

\author{Mikhail V. Medvedev }
\address{Canadian Institute for Theoretical Astrophysics,
University of Toronto,\\ Toronto, Ontario, M5S 3H8, Canada\\
Harvard-Smithsonian Center for Astrophysics, 60 Garden Street,
Cambridge, MA 02138}

\maketitle

\begin{abstract}
We present analytical and numerical solutions which describe a hot,
viscous, two-temperature accretion flow onto a neutron star or any
other compact star with a surface.  We assume Coulomb coupling between
the protons and electrons, and free-free cooling from the electrons.
Outside a thin boundary layer, where the accretion flow meets the
star, we show that there is an extended settling region which is
well-described by two self-similar solutions: (1) a two-temperature
solution which is valid in an inner zone $r\le10^{2.5}$ ($r$ is in
Schwarzchild units), and (2) a one-temperature solution at larger
radii.  In both zones, $\rho\propto r^{-2},\ \Omega\propto r^{-3/2},\
v\propto r^0,\ T_p\propto r^{-1}$; in the two-temperature zone,
$T_e\propto r^{-1/2}$.  The luminosity of the settling zone arises
from the rotational energy of the star as the star is braked by
viscosity; hence the luminosity is independent of $\dot M$.  The
settling solution is convectively and viscously stable and is unlikely
to have strong winds or outflows.  The flow is thermally unstable, but
the instability may be stabilized by thermal conduction.  The settling
solution described here is not advection-dominated, and is thus
different from the self-similar ADAF found around black holes.  When
the spin of the star is small enough, however, the present solution
transforms smoothly to a (settling) ADAF.
\end{abstract}

\section*{Introduction}

At mass accretion rates less than a few per cent of the Eddington
rate, black holes (BHs) and neutron stars (NSs) are believed to accrete
via a hot, two-temperature, radiatively inefficient, quasi-spherical,
advection-dominated accretion flow, or ADAF \cite{NY94,NY95}. 
While the properties of BH ADAFs are quite well known, hot flows
onto NSs have not been investigated. Their properties, such as the luminosity,
spectra, torque applied to a central object, etc., are expected
to be different from the BH ADAFs because a NS has a surface 
while a BH has an event horizon \cite{NY95,NGMc97}.
Here we discuss  the structure of a hot accretion flow around
a NS \cite{MN01}. We do not attempt a detailed
analysis of the boundary layer region near the NS surface.

\section*{Self-Similar Settling Solution }

\begin{figure}[b!] 
\centerline{\epsfig{file=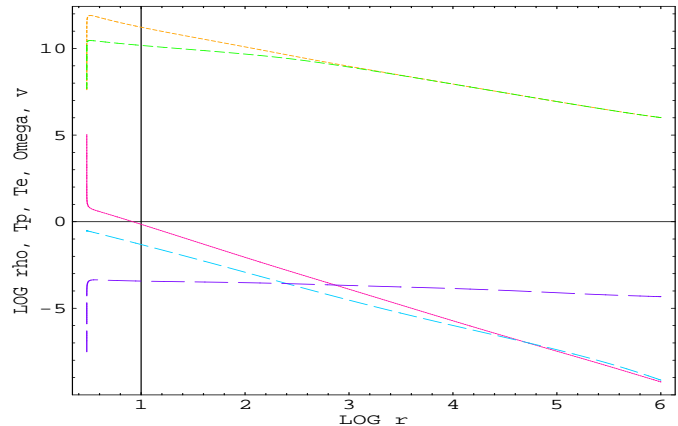,height=2.25in,width=3.5in}}
\vspace{10pt}
\caption{The radial profiles of density ({\em solid} curve), proton and electron
temperatures (two {\em dotted} curves), angular velosity ({\em dashed} curve),
and radial velosity ({\em long-dashed} curve).}
\label{fig1}
\end{figure}

We consider a steady, rotating, axisymmetric, quasi-spherical,
two-temperature accretion flow onto a star with a surface, and we use the
height-integrated form of the viscous hydrodynamic equations. We assume
the Shakura-Sunyaev-type viscosity parametrized by dimensionless $\alpha$.
We assume viscous heating of protons, Bremsstrahlung cooling of electrons
and Coulomb energy transfer from the protons to the electrons. We neglect 
thermal conductivity and Comptonization. In the inner zone $r<10^{2.5}$
($r$ is in Schwarzchild units, $R_S=2GM/c^2$), the flow is two-temperature
with the density, proton and electron temperatures, angular and
radial velocities scalings as 
\beq
\rho=\rho_0\,r^{-2},\quad T_p=T_{p0}\,r^{-1},\quad 
T_e=T_{e0}\,r^{-1/2}, \quad \Omega=\Omega_0\,r^{-3/2},\quad v=v_0\,r^0, 
\label{sss}
\eeq
where $\rho_0,\ T_{p0},\ T_{e0}, \Omega_0, v_0$ are functions
of $M,\ \alpha$ and the star spin $s={\Omega_*/\Omega_K(R_*)}$, and
$\Omega_K(R)=(GM/R^3)^{1/2}$ is the Keplerian angular velocity.
In the outer zone $r>10^{2.5}$, we have $T_e=T_p\propto r^{-1}$ and the
same other scalings. This self-similar solution is valid for the
part of the flow below the radius $r_s$ related to the mass accretion
rate $\dot m$ (in Eddington units, $\dot M_{\rm Edd}=1.4\times10^{18}m
\textrm{ g/s}$, and here $m=M/M_{\sun}$):
\beq
\dot m<2.2\times10^{-3}\alpha_{0.1}^2s_{0.3}^2r_{s,3}^{-1/2},
\label{out-constr}
\eeq 
where $r_{s,3}=r_s/10^3$,\ $\alpha_{0.1}=\alpha/0.1$, etc..
The numerical solution of the hydrodynamic equations with appropriate 
inner and outer boundary conditions is represented in Figure \ref{fig1}.
It is in excellent agreement with the self-similar solition (\ref{sss}).

\section*{Properties of the self-similar solution}

{\it Spin-Up/Spin-Down of the Neutron Star ---}
The angular momentum flux in the flow, $\dot J$, is negative which implies 
that the accretion flow removes angular momentum from the star and spins 
it down. This behavior is quite different from that seen in thin disks 
\cite{PN91,Paczynski91}, where for most choices of the stellar spin
parameter $s$, the accretion disk spins up the star with a torque
$\dot J_{\rm thin}\approx\dot M\Omega_K(R_*)R_*^2$.
In contrast, for the self-similar solution derived here, the torque is
negative for nearly all values of $s$. Moreover, $\dot J$ is independent 
of $\dot M$. Equivalently, the dimensionless torque, 
$j=\dot J/\dot J_{\rm thin}$, which is $\sim1$ under most 
conditions for a thin disk, here takes on the value 
$
j\simeq-43{\dot m}^{-1}{\alpha^2}s^3 \left(1-s^2\right)^{3/2} .
$
This torque spins down the NS as $s={s_0}/{\sqrt{1+t/\tau}}$, where the
spin-down time is
\beq
\tau\simeq2\times10^{8}s_{0.1}^{-2}\alpha_{0.1}^{-2}
\left({R_m}/{R_*}\right)^{-3/2}\textrm{ yr  ~~or~~  }
{\dot P_*}/{P_*^2}
\simeq2.7\times10^{-12}m_{1.4}^{-1}\alpha_{0.3}^2s_{0.5}^3\textrm{ s}^{-2},
\eeq
which is in excellent agrement with observational spin-down rates of some
X-ray pulsars \cite{YWV97} (here $R_m$ is the magnetospheric radius). 
Note, the spin-down rate is independent of
$\dot M$!

{\it Luminosity and Spectrum ---}
The total luminosity has two contributions: from the settling flow and 
from the boundary layer:
\bea
L_{SS}&\simeq&
6.2\times10^{34}mr_3^{-1}\dot m_{-2}s_{0.1}^2
+8.9\times10^{33}mr_3^{-1}\alpha_{0.1}^2s_{0.1}^4 
\textrm{ ergs/s}, \nonumber\\
L_{BL}&\simeq& 1.7\times10^{36}mr_3^{-1}\dot m_{-2}\textrm{ ergs/s} .
\label{L}
\eea
Note that for sufficiently low $\dot m$, the luminosity is independent
of $\dot m$ and is dominated by the settling flow. Below the radius
$r_c\sim45\alpha_{0.1}^{1/2}s_{0.1}$ Comptonization is significant
(although optical depth is always smaller than unity); therefore
the self-similar solution (\ref{sss}) is not accurate.
The observed spectrum 
from the settling flow (assuming free-free emission) is calculated to be:
 \beq
\nu L_\nu\simeq1.7\times10^{31}m\alpha_{0.1}^2s_{0.1}^4
\left({h\nu}/[{3\textrm{ keV}}]\right)\ \textrm{ ergs/s}.
\label{spec}
\eeq
which is sufficiently accurate up to 
$h\nu\sim kT_e(r_c)\sim400\alpha_{0.1}^{-1/4}s_{0.1}^{-1/2}\textrm{ keV}$.

{\it Stability of the flow ---}
We demonstrate that the settling flow is convectively stable
and may not have strong winds and outflows (the Bernoulli number 
is negative) if the adiabatic index satisfies
\beq 
\gamma>(3/2)\left(1-s^2/2\right)/\left(1-s^2/4\right)\sim1.5 .
\label{b<0}
\eeq 
The flow is thermally unstable because it is cooling-dominated by
free-free emission. Stabilization by thermal conduction is studied elsewhere.

\end{document}